\documentclass{ws-procs975x65}            

\newcommand{\beq}{\begin{equation}}
\newcommand{\enq}{\end{equation}}
\newcommand{\be}{\begin{eqnarray}}
\newcommand{\ee}{\end{eqnarray}}

\begin{document}
\title{B-mode in CMB polarization. What's that and why it is interesting.}

\author{A. D. Dolgov}

\address{
Dipartimento di Fisica, Universit`a degli Studi di Ferrara,  Polo Scientico e Tecnologico - Edicio C, Via Saragat 1, 44122 Ferrara, Italy
\\
Laboratory of Cosmology and Elementary Particles, Novosibirsk State University, Pirogov street 2, 630090 Novosibirsk, Russia \\
Institute of Theoretical and Experimental Physics, Bolshaya Cheremushkinskaya ul. 25, 113259 Moscow, Russia
\\
E-mail:  dolgov@fe.infn.it}

\begin{abstract}
Generation of the B-mode of CMB polarization by background of relic gravitational wave is discussed in connection with the BICEP2
measurements. Description of the polarization maps in terms of the eigenvectors of the polarization matrix is considered.

 is emphasized.    
\end{abstract}

\keywords{CMB polarization, B-mode, primordial gravitational waves}

\bodymatter

\section{Introduction}
Approximately three months ago striking news appeared in mass media  and arXive~\cite{bicep2} that the 
BICEP2 (Background Imaging of Cosmic Extragalactic Polarization) group announced the 
discovery  of a specific  form of polarization of the Cosmic Microwave Background 
Radiation, the so called B-mode, which may be an imprint of
very long gravitational waves and a proof of the cosmological inflation. 
In the recent revised version, which appeared s few days ago, at June, 23, the statement about the discovery
was somewhat milder. As it is written in the revised abstract "Accounting for the contribution of foreground dust will shift this value 
downward by an amount  which will be better constrained with upcoming data sets."

\begin{figure}
\begin{center}
\includegraphics[height=0.4\textheight]{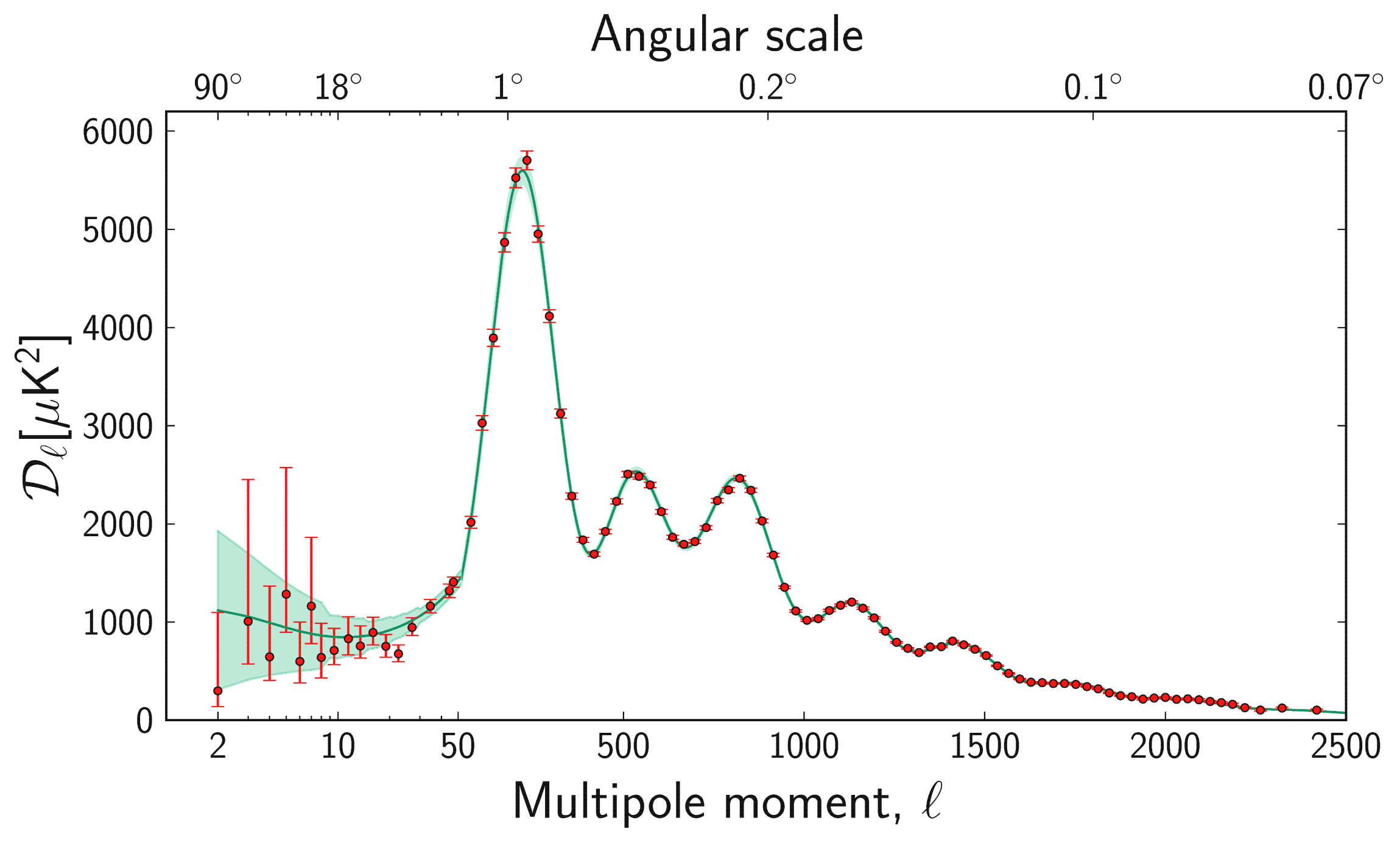}
              \caption{
              Spectrum of angular fluctuations of CMB measured by Planck.  Green (shaded) area shows  the cosmic variance.
                        } \label{fig-cmb-obs}
\end{center}
\end{figure}

The Cosmic Microwave Background (Radiation) or CMB(R) is the electromagnetic radiation  with the wave length around  ${\lambda \sim 0.1}$ cm,
with the perfect Planck (Bose-Einstein) spectrum: 
\be
f_\gamma = \frac{1}{ e^{\omega/T} -1}.
\label{f-gamma}
\ee
Possible chemical potential, which would indicate a deviation from the perfect thermal equilibrium is strongly bounded by
{${\mu/T < 10^{-4}}$; a similar limit is valid for the so-called y-distortion.} The temperature is quite accurately measured. 
{${ T = 2.725 \pm 0.002 }$K} all over the sky with very small, but non-zero, angular temperature fluctuations ${\delta T /T}$ below ${ 10^{-4}}$.
These fluctuations  are depicted in fig.~\ref{fig-cmb-obs}.
The angular inhomogeneities of the CMB temperature present a snapshot of
the universe after hydrogen recombination, when the universe was about 300,000 years old or at
redshift $ z_{rec} = 1100$. The angular fluctuation spectrum depends upon the cosmological parameters.
chemical content, and the expansion history of the universe.
An analysis of the spectrum of  ${\delta T}$ is one of the most precise  "tools"~to measure 
these parameters,  in particular: 
 ${ H,\,\, \Omega_{tot},\,\, \Omega_{DM},\,\, \Omega_{DE}, \Omega_B,\,\, n.}$ 
and even to detect primordial GWs. Here $\Omega_a =\rho_a/\rho_c$ is the cosmological fraction of the energy density 
of type $a$ matter and $n$ is
a power the spectrum of the density perturbations as a function of their wave length, see below, 
eqs. (\ref{h-perturb}) and (\ref{delta-rho-flat}).

The data present a strong evidence in favor  of inflation even in absence of registered long gravitational waves. Firstly, the position of
the first peak proofs that the universe is geometrically flat, as predicted by inflation, and, secondly, the measured value of $n$,
which is a little below unity,  is in beautiful agreement with the mechanism of inflationary generation of  density perturbations~\cite{mukh-chib}.

\section{A few words about inflation}

At this stage and in this audience it is proper to say a few words about inflationary cosmology. It is probably the greatest breakthrough in cosmology at the
second part of the XX century. Inflationary hypothesis~\cite{infl} explains in a simple and unique way all fundamental mysterious features of the Friedmann
cosmology: \\ 
1. The origin of the cosmological expansion (induced by primordial antigravity).\\
2. Isotropy: in the Friedmann cosmology two points separated by more than 1$\bm{^o}$ never knew
about each other, but the temperature of CMB is practically the same over all the sky. \\
{3. Homogeneity of the universe at large scales.}\\
4. Flat Euclidean geometry with the predicted precision of ${\sim 10^{-4}}$ precision; the observations confirm that  at the level of a few${\times 10^{-2}}$ .\\
5. Density perturbations with almost flat, Harrison-Zeldovich (HZ) spectrum~\cite{hz}.

The last, fifth, point deserves more attention. The spectrum of density perturbations is usually parametrized 
in  a power law form: 
\be 
h^2_k\sim k^{-3+(n-1)}, 
\label{h-perturb}
\ee
where ${h_k}$ is the Fourier transform of the correlator of the scalar metric perturbation
${ G(x-y)=\langle h(x) h(y)\rangle} $. Since metric is a dimensionless quantity the correlator should also be dimensionless.
If we assume that there is no dimensional parameter, i.e. the spectrum is scale-free, we come to the HZ-spectrum, which
corresponds to  ${n=1}$ and  ${ h_k^2 \sim {1}/{k^3}}$, and leads to dimensionless $G \sim \int d^3k /k^3$ without  
necessity of any internal scale.

Density perturbations are obtained from the metric perturbations through the Poisson equation, $\Delta h \sim \rho$ and hence:
 \be 
\left({\delta\rho}/{\rho}\right)^2_k  \sim k^4  h_k^2 \sim k^n.
\label{delta-rho-flat} 
\ee
Inflation predicts ${ n}$ slightly smaller than unity. According to observations  ${n= 0.96}$. The theoretical value 
is model dependent and well fits ${R^2}$-inflation, proposed in the first paper in ref.~\cite{infl}.

Can we conclude on the basis of these data that inflation is already an "experimental" fact? 
I think yes, but some some people still think that 
the final proof would be an observation of very long GWs produced at inflationary stage. However, the intensity 
of these GWs depends on the mechanism of inflation and on the underlying 
cosmological model and may be very small. Surely observations of GWs would prove 
existence of inflation beyond any doubts, but the absence of GWs would not disprove inflation.

\section{Gravitational waves}

GWs were predicted by Einstein, but he himself had long alternating feeling  about their
existence. Now it is established that 
metric perturbations can propagate with the speed of light and carry energy. They satisfy the wave equation:
\be 
(\partial^2_t  - \Delta) \psi^i_j = 16 \pi G_N T^i_j ,
\label{GW-eqn}
\ee
where the source, ${T^i_j}$, is the quadrupole component of energy-momentum tensor
of matter and the function $\psi$ is expressed through the metric perturbations as: ${ \psi^i_j = h_i^j - \delta_i^j h/2 }$.
In contemporary universe GWs can be produced in catastrophic stellar processes, by compact star binaries, and more... 

We are speaking here, however, about  cosmological gravitational waves. The
Parker theorem~\cite{parker} forbids production of massless particles in the conformally flat space-time, in particular, in
the Friedmann cosmology due to conformal invariance of wave equations for massless particles.
However,  it was found by Grischuk~\cite{grischuk} that gravitons are not
conformally invariant and time dependent cosmological gravitational field could produce GWs.
The calculation of the  
intensity of GWs produced at the De Sitter (inflationary) metric was pioneered in ref.~\cite{gw-infl}. It was  
shown that long gravitational waves with rather high intensity, $ \Omega_{GW} \sim H_{inf}^2 / m_{Pl}^2$, (here $ H_{inf} $ is the
Hubble parameter during inflation) 
could be generated. More precisely, elementary particles, including quanta of GWs,
were produced in the course of  the {transition} from DS to Friedmann regime. Inflationary
GW production is quite similar to the generation of density perturbations with the difference that the former are scalar perturbations, while
GWs are tensors ones.

\begin{figure}
\begin{center}
\includegraphics[height=0.6\textheight]{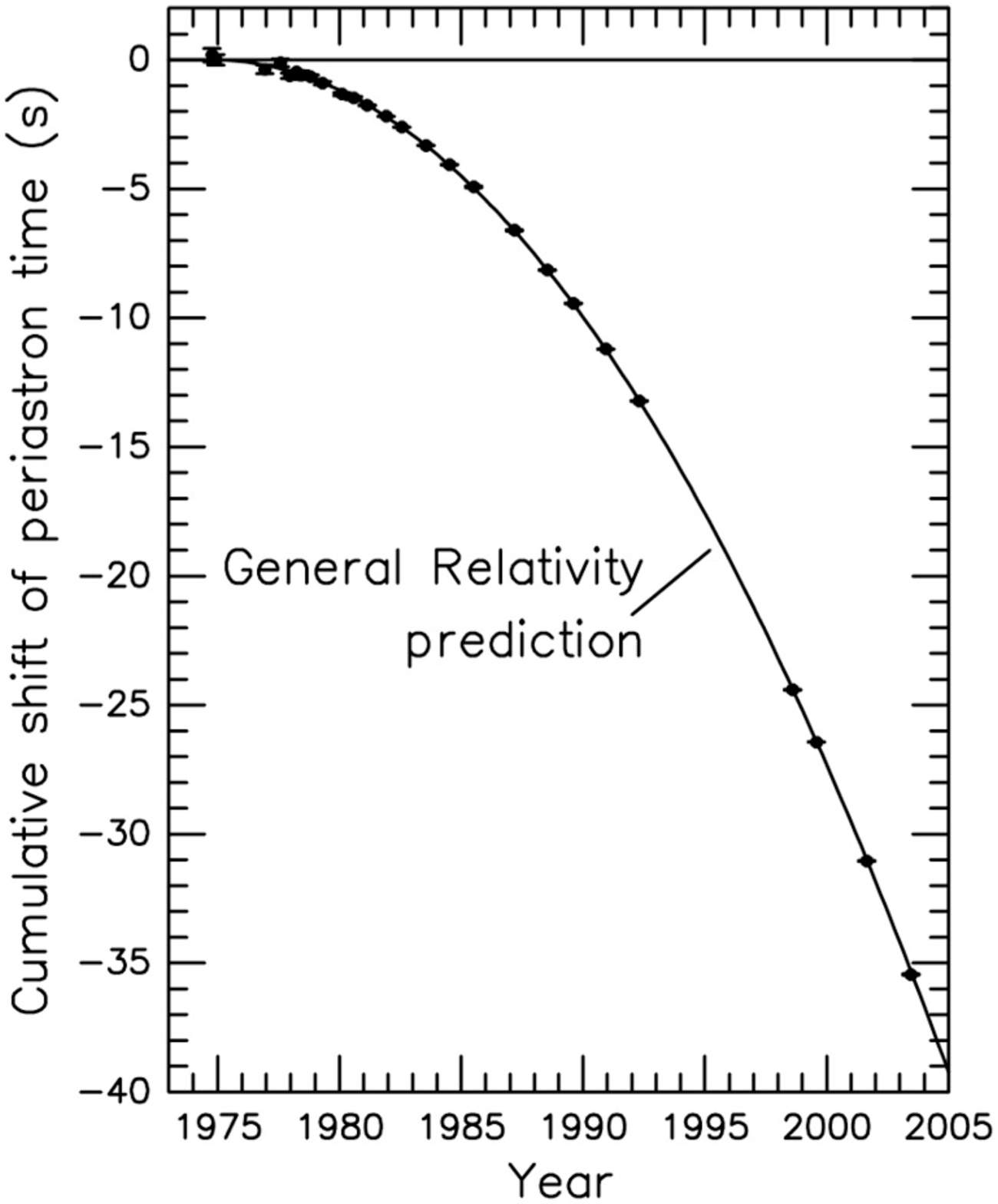}
\caption{{ Decrease of a double pulsar orbital period.}}
            \label{f-pulsar}
             \end{center}
\end{figure}

Up to the present time GWs are not yet observed anywhere, though an indirect evidence in favor of their existence by observations of the 
energy loss in double pulsar systems is very strong, see fig.~\ref{f-pulsar}.

\begin{figure}
\begin{center}
\includegraphics[height=0.5\textheight]{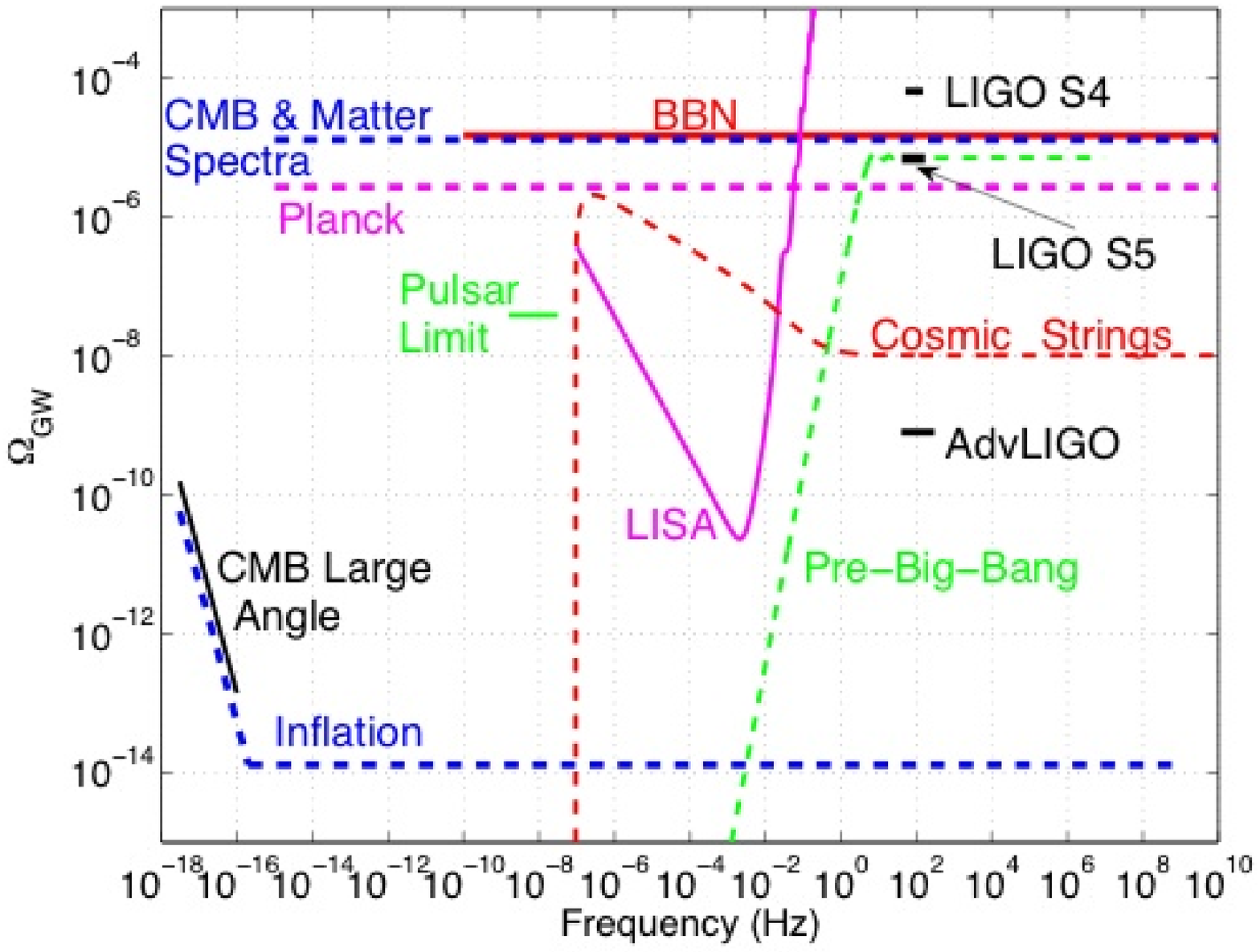}
           \caption{${ \log [h_0^2\Omega_{GW}}$(f)] vs. ${ \log (f [Hz])}$ for different models of 
production of stochastic background of GWs as given by Abbot et al (LIGO and VIRGO Coll.)}
\label{f-limits}
             \end{center}
\end{figure}

There are a few detectors dedicated to the search GWs in the universe, among them are 
LIGO = Laser Interferometer Gravitational Wave Observatory
and LISA = Laser Interferometer Space Antenna (in project). The latter may be  
sensitive to the inflationary GWs. It is  able to measure the relative displacements with a 
resolution of 20 picometers over  5 million kilometers, 
strain sensitivity better than 1 part in ${10^{20}}$.
For comparison, inflation predicts the strain ${ h \sim 10^{-18} (10^{-10} \, Hz) / f)^2 }$.
Sensitivity of different detectors and some physical phenomena in cosmology to the cosmic background of 
GWs is shown in fig.~\ref{f-limits}.

Inflation predict GWs with the flat spectrum i.e.  with the intensity independent on their frequency. However, we see
in fig~\ref{f-limits} that $\Omega_{GW}$ drops down at very small frequency and then remains constant in all frequency range.
The reason for that is the following. The intensity of GWs produced at the end of inflation is given by:
\be 
\Omega_{GW}^{(infl)} \sim \frac{H_I^2}{m_{Pl}^2} \approx
\frac{m^2_\phi \phi^2}{m_{Pl}^4} \sim \frac{m_\phi^2}{m_{Pl}^2},
\label{Omega-GW}
\ee
where $\phi$ is the inflaton field and $m_\phi$ is its mass.
The last equality is fulfilled because at the end of inflaton the field $\phi \approx m_{Pl}$. 
Small magnitude of the primordial cosmological density perturbations,   $\delta \rho /\rho \sim 10^{-4}$  demands 
${  m_\phi / m_{Pl} \sim 10^{-5} }$. That's why  ${{\Omega_{GW}^{(infl)} \sim 10^{-10}}}$ at very low frequency.
 After transition to the cosmological MD-stage at the redshift $z=10^4$ the energy density of GWs is redshifted to
 the present time by 4 orders of magnitude, so
 ${\Omega_{GW}}$ dropped by the same amount down to ${10^{-14}}$. This occurred with 
 the waves with the lengths shorter than horizon at the matter-radiation equilibrium epoch, while longer waves 
 remained undisturbed.

\section{CMB polarization. Different ways of  mapping}

With  the presently existing sensitivity of GW  detectors the perspectives of direct registration of the inflationary GWs are not particularly 
bright but the chances to observe GWs through the B-mode of the CMB polarization are feasible. 
Polarization of CMB photons is created by scattering  of initially unpolarized photons on (unpolarized) electrons in cosmic plasma
because the probabiity of scattering depends, in particular, on polarization of the scattered photon.
There are two possible types of polarization: scalar (E-mode) and pseudoscalar (B-mode).
Without GWs only E-mode is induced. In fact not only GWs can induce the B-mode but also
GWs, as well as intergalactic magnetic fields, CMB lensing, scalar perturbations in the second order, and last but not
the least, rotating dust. Some more detail can be found in lecture~\cite{ad-paris}.

Polarization of electromagnetic radiation is described by the polarization density matrix:
\be 
\rho_{ij} =\langle E_i E^*_j \rangle,
\label{rho-ij}
\ee
where $E_i$ is the vector of electric field of the electromagnetic wave.
$\rho_{ij}$ is a 2nd rank tensor in two dimensional  space orthogonal to the photon propagation. We take it  as ${(x,y)}$-space
if photon propagates along  $z$-direction. Such tensor has two well known algebraic invariants,
trace, which is equal (or proportional) to the intensity of radiation: 
\be 
T  = \delta_{ij}\rho_{ij} = |E_x|^2 + |E_y|^2 
\label{T}
\ee
and helicity: 
\be
 V = \epsilon_{ij} \rho_{ij} .
\label{V}
\ee
The density matrix can be expanded in terms of the full set of $2\times 2$ matrices:
\be 
\rho_{ij} = T \left( {I}/{2} + \xi_k \sigma_k \right),
\label{stokes}
\ee
where ${I}$ is the unit matrix, ${\sigma_k}$ ($k=1,2,3$) are the Pauli matrices, The coefficients 
$\xi_i$ are called the Stokes parameters. At rotation around $z$-axis $\xi_1$ and $\xi_3$ are transformed
through each other and with a proper choice of the coordinates it is possible to eliminate $\xi_1$,
$\xi_1 =0$, while $\xi_2$ is invariant. It describes the photon helicity. If we neglect parity nonconservation,
which is usually induced by weak interactions, the an initially zero photon helicity would remain zero and one
can out $\xi_2 =0$ as well.

If $\xi_2 =0$, the traceless part of the polarization matrix is described by  two functions, ${Q}$ and ${U}$: 
\be \bar\rho =
\left[ \begin{array}{cc}
Q & U \\
U & -Q  
\end{array}\right]
\label{rho-bar}
\ee

Evidently the polarization should vanish in homogeneous and isotropic  world.
The concrete expression for the polarization due to Thomson scattering can be obtained by
integration of the scattering probability for different polarization 
over angles 
with rotation around ${z}$ to common coordinate system:
\be
Q -i U =\frac{\sigma_T}{\sigma_N}\,\int d\omega \sin^2\theta \exp[2i\phi]
T'\left(\theta,\phi \right)
\label{Q-iU}
\ee
where $\sigma_T$ is the Thomson cross-section and ${\sigma_N}$ is a normalization area.
One can see that the polarization is proportional to the quadrupole moment of radiation.

There are two more (now differential) invariants of the polarization matrix:
scalar: ${ S=\partial_i \partial_j \rho_{ij}}$ (E-mode) and
pseudoscalar:  ${ P=\epsilon_{ik}\partial_i \partial_j \rho_{jk}}$ (B-mode).
For purely scalar perturbations the only way to write the polarization matrix is:
\be 
\bar\rho_{ij} = \left(2 \partial_i \partial_j -\delta_{ij} \partial^2\right)\Psi ,
\label{rho-scalar}
\ee
where $\Psi$ is a scalar function.
Correspondingly $ P=0. $  Non-zero P is an indication for something
extra beyond scalar perturbations.

Apart from scalars there could be: \\
{1. Vector perturbations, e.g. (intergalactic) magnetic fields,}
\be 
\bar\rho_{ij} = \partial_i V_j - \partial_j V_i ,\,\,\,\,\,
P = \epsilon_{ij}\partial^2 \partial_i V_j
\label{rho-vect}
\ee
(created by scattering in magnetized medium).\\
2. Tensor perturbations, e.g. GWs,
\be 
\bar\rho_{ij} \sim  \partial^{-2} ( \partial_i h_{3j} - \partial_j h_{3i}),\,\,\,\,\,
P \sim \epsilon_{ik}\partial_i h_{3k}.
\label{rho-GW}
\ee
{{ 3. Second order scalar perturbations,}} 
\be 
\bar\rho_{ij} \sim \partial_i \Psi_1 \partial_j \Psi_2 - 
\partial_i \Psi_2\ \partial_j \Psi_1, \,\,\,\,\,
P = \epsilon_{ik} \partial_i ( \Delta \Psi_1 \partial_k \Psi_2 - \Delta \Psi_2 \partial_k \Psi_1 )
\label{rho-2}
\ee
(e.g. for ${\Psi_2 =\partial_t \Psi_1}$).

All such types of perturbations result in $P\neq 0$ and thus they can create B-mode of polarization.

Usually the ``direction'' of polarization is characterized by the so called ``vector'' given by two components:
\be
v = (Q,U) ,
\label{vector-QU}
\ee
though ${v}$ is not a vector but a mixture of 2nd rank tensor components. 
As a result of that the polarization map in terms of $(Q,U)$ changes under rotation, while it should not 
if $v$ would be a real vector. Because of that it would be proper to present the polarization map in terms
of real vectors, which are the eigenvectors of the polarization matrix ${\rho_{ij}}$. This would allow to make 
an invariant (covariant) description with respect to the choice of coordinates.

\begin{figure}
\begin{center}
\includegraphics[height=0.2\textheight]{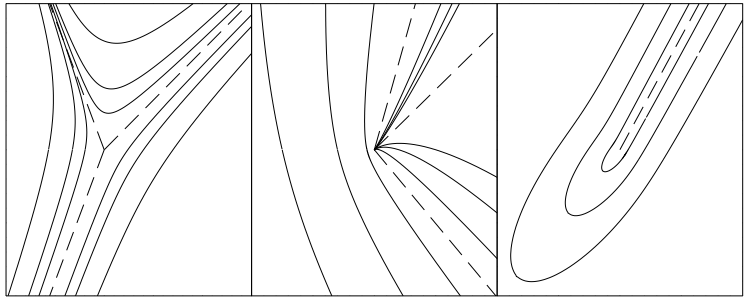}
           \caption{Flux lines for three different types of singular points: saddle, beak, and  comet.
Dashed lines show peculiar solutions (separatrix).}
\label{f-sing-eigen}
             \end{center}
\end{figure}

The singularity points of ${v}$ are the usual ones: saddles, foci, knots, which are
well known from the  classical analysis
of dynamical systems. It is troublesome that different singularity points transform into each other under rotation!
On the other hand, the singularity points in terms of real vectors do not depend on the coordinate choice, as it can be 
seen for the case of  ejgenvectors of $\rho$. It is interesring that the eigenvectors possess new types
of singularities, because of non-analytic properties of the eigenvectors, at the points where they 
are zero, see fig.~\ref{f-sing-eigen}.

The suggestion to use eigenvectors of polarization matrix and comparison with the traditional method are made in our 
works~\cite{ddnn}.

Let us stress, in reply to an often asked question, that the polarization state in a fixed coordinate point is described by
a single number, the value of $\xi_3$ Stokes parameter. The difference between S(E) and P(B) modes can be observed 
by a distribution of polarization at different points (i.e. on the whole map). Recall that  S(E) and P(B) modes are related to 
the differential invariants, see their definitions above eq.~(\ref{rho-scalar}).   Because of that, in particular, the 
directions of $v$"vector" corresponding to scalar (E) mode 
and the gradient of $v$-"vector" corresponding to pseudoscalar (B) mode differs by 45$\bm{^o}$.

According to the BICEP2 measurements~\cite{bicep2} the B-mode is observed with quite large magnitude which corresponds 
to the ratio of tensor to scalar perturbations equal to:  {${ r =0.20^{+0.07}_{-0.05}}$,} while
{r = 0 is disfavored at ${7.0\sigma}$.}
Subtracting the best available estimate for foreground dust modifies the likelihood slightly, so that $r = 0$
is disfavored at ${5.9\sigma}$. 

On the other  hand the Planck estimate of foreground created by dust is much higher than it was
estimated by BICEP. Accordingly the Planck results allow to put only an upper bound $ r < 0.11 $~\cite{planck-r}.

\section{Conclusion.}

The new result ob BICEP2 is intriguing and stimulating for further research, both in observation and theory.
An analysis of other foregrounds is in order and independent confirmations are necessary.
It seems very desirable to make measurements at different frequency bands. 
Description of the polarization field in invariant eigenvectors of the polarization matrix can be helpful.

I acknowledge the support by the grant of the Russian Federation government 11.G34.31.0047.


\end{document}